\documentclass{aa}

\def\arcsec              {$^{\prime\prime}$}

\usepackage{graphics}
\usepackage{latexsym}
\begin{document}
  \thesaurus{(03.20.1; 11.17.4 Q~2237+0305; 12.07.1)}
  \title{Three   photometric methods tested  on ground-based data of
  Q~2237+0305\thanks{Based on observations obtained at NOT, La Palma.}}
  \author{I. Burud\inst{1,2}, R. Stabell\inst{2,3}, P. Magain\inst{1}\fnmsep\thanks{Ma\^
  {\i}tre de Recherches au Fonds National Belge de la Recherche Scientifique}, 
  F. Courbin\inst{1,4}, R. {\O}stensen\inst{1,5}, S. Refsdal\inst{3,6}, M. Remy
  \inst{1}, J. Teuber\inst{3}}
  \offprints{I. Burud, Li{\`e}ge address.   Email: burud@astro.ulg.ac.be}
  \institute{Institut d'Astrophysique, Universit{\'e} de Li{\`e}ge,
  Avenue de Cointe 5, B--4000 Li{\`e}ge, Belgium
  \and
  Institute of Theoretical Astrophysics, University of Oslo,
  Pb. 1029 Blindern, 0315 Oslo, Norway
  \and
  Centre for Advanced Study, Drammensveien 78, N-0271 Oslo, Norway
  \and 
  URA 173 CNRS-DAEC, Observatoire de Paris,
  F--92195 Meudon Principal C{\'e}dex, France
  \and
  Department of Physics, University of Troms{\o}, N-9037 Troms{\o}, Norway
  \and
  Hamburger Sternwarte, Gojenbergsweg 112, D-21029 Hamburg, Germany}

\date{Receiveved 01 July, 1998; Accepted 25 August ,1998}

\titlerunning{Photometry of the Einstein Cross}
\authorrunning{I. Burud et al.}
 
\maketitle

\begin{abstract}

The Einstein Cross, Q~2237+0305, has  been photometrically observed in
four  bands on two   successive nights  at  NOT  (La Palma,  Spain) in
October 1995.  Three independent algorithms  have been used to analyse
the  data:  an  automatic   image  decomposition technique,   a  CLEAN
algorithm  and  the new MCS deconvolution   code.  The photometric and
astrometric results obtained with the three methods are presented.  No
photometric variations   were   found  in  the  four  quasar   images.
Comparison of the photometry from the three techniques shows that both
systematic and random errors affect  each method.  When the seeing  is
worse than    $1\farcs0$,  the   errors from  the     automatic  image
decomposition   technique and  the  Clean algorithm   tend to be large
(0.04-0.1   magnitudes)  while   the deconvolution  code   still gives
accurate  results (1$\sigma$  error below  0.04) even  for frames with
seeing as bad as $1\farcs7$.

Reddening  is observed   in the quasar  images    and is found  to  be
compatible with either extinction   from the lensing galaxy or  colour
dependent microlensing.

The  photometric accuracy depends on  the  light distribution used  to
model the lensing galaxy.    In particular, using a numerical   galaxy
model,  as done with the MCS  algorithm, makes the  method less seeing
dependent.   Another advantage of  using   a numerical model is   that
eventual non-homogeneous structures in the galaxy can be modeled.

Finally, we propose an observational strategy for a future photometric
monitoring of the Einstein Cross.
\end{abstract}

\section{Introduction}

The gravitational  lens   system Q~2237+0305,  known  as the  Einstein
Cross,  is  one of  the most  promising   objects to observe intensity
variations due to    microlensing   of quasar images.     The  object,
discovered by Huchra et  al.  (\cite{huchra}), consists of four images
of the same quasar at $z=1.69$ lensed by a foreground spiral galaxy at
$z=0.04$.  The maximum  angular separation of  the components is about
$1\farcs8$.  Due to the proximity  of the lens  and the high degree of
symmetry of the system, the time delays  between the images are of the
order of  one day.  Intrinsic  variations of the source therefore show
up almost simultaneously in all  four quasar images, hence making them
easy to  distinguish  from microlensing effects.   This  system with a
low-redshift lens is an  ideal case for studying microlensing effects,
since the light  paths to the  different  QSO images pass through  the
bulge of the galaxy, hence increasing the probability of gravitational
influence by single  stars  (Chang \& Refsdal  \cite{chang}, Paczynski
\cite{pac},  Kayser    et al.     \cite{kaysera}, Kayser   \&  Refsdal
\cite{kayserb}).  In addition   we note that the  angular  size of the
Einstein ring for  a given lens mass is  larger than  for systems with
higher lens redshifts.

However, detecting faint  intensity variations in multiply imaged QSOs
requires very  accurate  photometry.  For most  gravitationally lensed
QSOs, this is not a straightforward task, Q~2237+0305 being one of the
most  complicated   cases.  Given the   blending  of  the  QSO images,
aperture photometry is excluded and profile  fitting photometry is not
trivial.  Moreover,  since the foreground galaxy  has a sharp nucleus
creating  a   non-uniform  and  fast-varying  background,    the light
distribution of the lensing galaxy has to be modeled carefully.

Several  methods and algorithms  have  been  developed to perform  PSF
photometry of   blended sources and   more  specific  codes have  been
developed  to treat the Einstein  Cross.  We compare in the following,
three  of these methods  (described  in section 3):  an  automatic PSF
fitting technique    (M.   Remy  \cite{remy},   hereafter  the Fitting
method),  an      interactive   CLEAN   algorithm   (R.    {\O}stensen
\cite{roya}), and a deconvolution method (P.   Magain, F.  Courbin
\& S.  Sohy \cite{magain}, hereafter MCS).

For this purpose, numerous images of the Einstein  Cross were taken on
two successive nights at  the  Nordic Optical  Telescope (NOT)  at  La
Palma, Canary Islands (Spain).  A homogeneous set of data taken over a
short  time scale with  a  good temporal sampling   was obtained.   No
physical intensity variations were likely to be detected in the object
during such a  short period so that the  data set is very well  suited
for performing photometric tests.  For  each  method, we measured  the
photometric robustness  with  respect  to  the  seeing  variations  in
optical $B,V,R,I$-bands.

If  we detected real variations  during the two  nights, we would have
had the  chance to  observe either  a high  amplification microlensing
event  (if the   variation had  occurred  in one  image   only)  or an
intrinsic  fluctuation in the  quasar itself, which  could have led to
the determination of the time delay for this system.

\section{Observations}

The observations took place at the NOT on the nights of October 10 and
11,  1995.  We used the CCD  camera BroCam  1  equipped with a thinned
backside  illuminated  TEK  1024   CCD with a   conversion  factor  of
$q=1.7e^{-}ADU^{-1}$,  a readout noise of  $6.5e^{-}$ and a pixel size
of $0\farcs176$.  Sequences  of exposures  were obtained  through  the
filters $B,V,R$ and $I$ (in  this order).  The  total number of frames
obtained of  the Einstein Cross in   each band, as  well  as the total
exposure time    and  the  mean    seeing   value are   summarized  in
Table~\ref{tab:log}.  The    seeing  of  the   frames   varied between
$0\farcs6$ and $1\farcs7$.

\footnotesize
\begin{table}[htbp]
\centering
\caption{Log of observations of the Einstein Cross from the $10^{th}$ and $11^{th}$ of October 1995. The first two lines show the total number of frames in each band. The next two lines give the total exposure time and the mean seeing value.}
  \begin{tabular}[t]{lcccc}
   \hline
    &\bf B & \bf V & \bf R & \bf I \\
   \hline
   \bf $1^{st}$ night & 6 & 9  & 8& 9   \\
   \bf $2^{nd}$ night & 7&9 &7 & 9\\
   \bf  total exp. time& 3400 s& 4100 s&3100 s &3750 s \\
   \bf Mean seeing & 0\farcs94&0\farcs94 & 0\farcs86& 0\farcs82\\   
   \hline 
\label{tab:log}
\end{tabular}
\end{table}
\normalsize

Bias subtraction, flat-field   correction (sky-flats)  and cosmic  ray
removal were applied to the raw data using the  ESO MIDAS routines. In
order to model  and subtract low frequency  sky  variations across the
frames, bi-quadratic polynomial surfaces were fitted to a well sampled
grid of empty regions in each individual frame.

\section{Photometric methods}

\subsection{The Fitting method}

A profile fitting method  has been developed  in the MIDAS environment
by M.   Remy (\cite{remy}) in   order  to obtain accurate  photometric
measurements of multiply imaged quasars.  This method has already been
applied  to other lensed systems,   (e.g., H1413+117, the  Cloverleaf,
{\O}stensen et al.   \cite{royc}).  Unlike   the Einstein Cross,   the
Cloverleaf system  does not suffer from the  contamination of a bright
and complicated  foreground   lens, and is  therefore   much easier to
study.

In  the  present case,   the magnitudes of    the quasar  images  were
determined   by  fitting simultaneously  numerical  PSFs  to the point
sources, and a de Vaucouleurs function  ($R^{-1/4}$) to the galaxy. An
additional Moffat profile was  also  fitted (still simultaneously)  in
order to model the point-like galaxy nucleus.  All the parameters were
adjusted  by  the program,   using   a  $\chi^{2}$ minimization.    The
intensity and positions for the four numerical PSFs and for the Moffat
profile were determined as well as the shape and rotational parameters
of the Moffat and the de Vaucouleurs model.

The technique was tested in different ways on  our images.  First, the
fit was performed with all the parameters free  on the whole data set.
From  these  results, we  calculated   the mean positions  of  the QSO
components and the  galaxy nucleus.  Then we  ran the program with the
position parameters fixed relative to   component A.  Furthermore,   a
galaxy model with  fixed  shape parameters was obtained  by extracting
the galaxy from the best result in each band. It was then convolved to
the respective seeing   of  each individual  frame.    Another fit was
performed with  a fixed  galaxy model centered  on  the  nucleus, i.e.
only  its intensity  was  left as  free parameter.   The  advantage of
freezing the relative positions of the QSO images and the galaxy model
is to minimize the number of free parameters in the fit.  On the other
hand, if the positions and  the galaxy model are  not accurate, we may
introduce systematic errors.  In the present case, the best $\chi^{2}$
fit was obtained with all the positions fixed relative to component A,
and a   de Vaucouleurs  profile with   fixed parameters.  Section  4.1
presents the results obtained in this way.

\subsection{A CLEAN algorithm}

A program for  CLEAN photometry of  overlapping point sources has been
developed by R.  {\O}stensen (\cite{roya}), and implemented using IDL.
This   program,  called XECClean, has   been  specifically designed to
perform photometry on  the Einstein Cross and has  been applied to the
NOT monitoring data  of the object  ({\O}stensen et al.  \cite{royb}).
As for the Fitting  method, the programme has   also been used  on the
monitoring data of H1413+117 ({\O}stensen et al.  \cite{royc}).

XECClean applies  a   semi-analytical PSF-profile  fitting   procedure
adapted  from  the   DAOPHOT package   (Stetson  \cite{stetson}),  and
``deconvolves'' the images using an interactive CLEAN algorithm (e.g.,
Teuber \cite{teuber}).  A PSF  is successively fitted and removed from
each QSO image until a first convergence of the parameters is reached.
Once the  point  sources are removed,  a  de Vaucouleurs profile  with
fixed shape parameters is centered on the galaxy nucleus and convolved
to  the seeing of   each frame. This model  is  fitted to  the lensing
galaxy and subtracted from the image.   Then the quasar components and
the  galaxy are iteratively restored, fitted  again, and removed until
convergence is reached.  This  is repeated as  many times as necessary
for obtaining satisfactory residuals for each frame.

\subsection{The MCS deconvolution algorithm}

A new  deconvolution   method has been    developed by Magain  et  al.
(\cite{magain}).   Contrary to  traditional  methods of deconvolution,
this algorithm allows not  only a significant  increase in the spatial
resolution of the images, but also to perform accurate photometric and
astrometric measurements on the deconvolved frames.

The algorithm is  based on the principle  that sampled  data cannot be
fully deconvolved without recovering  Fourier frequencies higher  than
the Nyquist frequency, and violate the sampling theorem.  A sampled
image  should therefore not  be deconvolved by  the total PSF but by a
narrower   function chosen so that  the  resolution of the deconvolved
image is compatible with the adopted sampling.

The image  is decomposed into a sum   of point sources  plus a diffuse
background.   The background is   constrained  to be  smoothed  on the
length scale of  the final resolution, chosen  by the user.   The best
model image is computed by  minimizing the  $\chi^2$ using a  modified
version of the conjugate gradient method (Press et al. \cite{press} ).
Positions and intensities of the point sources as well as the image of
the  deconvolved background are  given as  output of the deconvolution
procedure.

\begin{figure}[p]
\begin{center}
\resizebox{7.5cm}{!}{\includegraphics{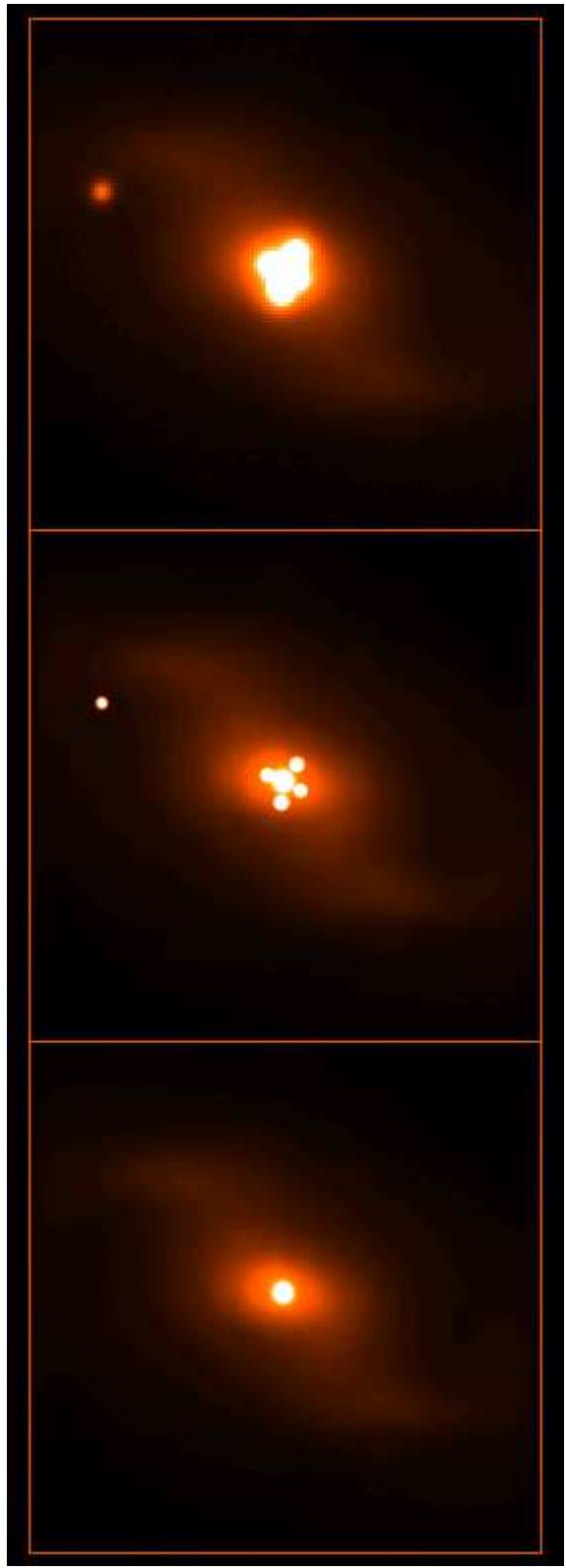}}
\caption{[{\it Top}]:  Stack of  12  R band  frames  with a  seeing of
$\sim$0\farcs8.   [{\it Middle}]:  Deconvolved   frame using  the  MCS
algorithm.  The  resolution  is    now 0\farcs26.    [{\it   Bottom}]:
Deconvolved   numerical   galaxy    model.    All   the  images    are
$\sim$22\arcsec\, on a side, North is up and East is to the left. }
\label{fig:rima}
\end{center}
\end{figure} 

Successful results have already been obtained on several gravitational
lens   systems    (e.g.,     Courbin  et    al.      \cite{courbina}a,
\cite{courbinb}b).  In the present case,  we  chose the pixel size  of
the deconvolved frame  to  be  half the one   in  the  original  data,
$0\farcs176/2=0\farcs088$.    Furthermore,   we  adopted  for      the
deconvolved point  source a Gaussian profile with  a FWHM of 3 (small)
pixels, which  allows us to reach  a  final resolution of $0\farcs26$.
The deconvolution  of our images  was performed in  two steps.  First,
the frames in a given filter were averaged, giving a deep image of the
object.  This  image was  deconvolved in order  to obtain  an accurate
numerical model  of the  lensing  galaxy.   Since the galaxy   profile
varies sharply in the  vicinity of the  nucleus, we applied a variable
smoothing   parameter $\lambda$ across the   field,  in order to avoid
local over- or  underfitting of the  data  (see Magain  et al.  1998).
Since the nucleus-shape is  close to that of a  point source, but  not
exactly, we added a  quasi-point-source described by a Moffat profile,
for the nucleus.  Its intensity,   position and shape parameters  were
all determined by the deconvolution program.

Fig.~\ref{fig:rima} presents the result obtained from the stack of our
R-band frames. The top and the middle panels show the observed and the
deconvolved  frame respectively, and  the   bottom panel displays  the
numerical galaxy model obtained   by  the algorithm.  Both the   point
sources (quasar components and foreground star) and the lensing galaxy
appear clearly without any deconvolution artifacts.  The separation of
the  point  sources  from  the  background  allows  one to  study  the
deflector  alone,  without  the  disturbing  light  from the  quasar's
images.  We  could in this  way determine an  accurate numerical model
for the   lensing galaxy,  that can   be used  for  future photometric
monitoring.

After a numerical model of the galaxy had  been obtained in each band,
it was  used  as  a fixed  background   for the deconvolution of   the
individual frames.  Only a  multiplicative factor and an additive term
were  applied in order to correct  for different exposure times (e.g.,
shutter effects) and varying  sky-levels.  The frames were deconvolved
both separately and  simultaneously.   The  advantage of  simultaneous
deconvolution is that the  solution is compatible  with all the images
considered.  The  intensities of  the point   sources  are allowed  to
converge   to different  values from  one  frame to    another so that
intrinsic variations  in    the object  can  be  detected,   while the
positions of the point sources are constrained by all the images used.

\section{Results}

\subsection{Photometric results}

Our   photometric magnitudes  are   calibrated   by  means  of   Yee's
(\cite{yee}) reference  star, using  the transformation  equations and
revised  values  given by  Corrigan    et al.  (\cite{corrigan}).   No
significant intensity variation was  detected  on the data  during the
two    nights of observation.  Therefore    a  mean magnitude could be
calculated  for each QSO image and  a  $1\sigma$ error for the results
obtained with  each method.  These values,  as well  as the calculated
error  on the mean  magnitude, are shown in Tables  2, 3, 4, and 5 for
the $B$,$V$, $R$, $I$ filters respectively.  The standard deviation of
the mean  $\sigma_{mean}$ was calculated  in order to give an estimate
of  the  error for  each  QSO image in  each filter  and is simply the
standard deviation  for all the frames divided  by the  square root of
the number of frames used.

\footnotesize
\begin{table}[p]
\centering
\caption{B-band photometric results: mean magnitude and standard deviations }
  \begin{tabular}[t]{lrrr}
   \hline
    &\bf Fitting & \bf Clean & \bf Deconv  \\
   \hline
   \bf A & $17.612\pm0.025$  & $17.626\pm0.033$  &  $17.655\pm0.029$  \\
      $\sigma_{mean}$    &       $\pm0.007$  & $\pm0.009$ & $\pm0.008$ \\
   \bf B & $17.754\pm0.040$  &  $17.741\pm0.030$ &  $17.754\pm0.023$\\
      $\sigma_{mean}$    & $\pm0.011$ & $\pm0.008$ & $\pm0.006$ \\
   \bf C & $18.937\pm0.076$  & $18.881\pm0.073$ &   $18.885\pm0.040$   \\
      $\sigma_{mean}$    & $\pm0.021$ & $\pm0.020$ & $\pm0.011$ \\
   \bf D &$19.155\pm0.034$  & $19.111\pm0.121$ &    $19.220\pm0.040$  \\
      $\sigma_{mean}$    & $\pm0.009$ & $\pm0.033$ & $0.011$ \\
   \hline 
\end{tabular}
\end{table}

\begin{table}[p]
\centering
\caption{V-band photometric results: mean magnitude and standard deviations. We have removed one of the frames with a FWHM=$1\farcs63$ in the results from
the Fitting and CLEAN algorithms.}
  \begin{tabular}[t]{lrrr}
   \hline
    &\bf Fitting & \bf Clean & \bf Deconv  \\
   \hline
   \bf A & $17.257\pm0.018$  & $17.278\pm0.019$  &  $17.276\pm0.016$  \\
      $\sigma_{mean}$ &       $\pm0.004$  & $\pm0.004$ & $\pm0.004$ \\
   \bf B & $17.414\pm0.014$  &  $17.428\pm0.022$ &  $17.425\pm0.015$\\
      $\sigma_{mean}$ &       $\pm0.003$  & $\pm0.009$ & $\pm0.004$ \\
   \bf C & $18.531\pm0.062$  & $18.389\pm0.028$ &   $18.415\pm0.035$   \\
      $\sigma_{mean}$ &       $\pm0.015$  & $\pm0.009$ & $\pm0.008$ \\
   \bf D &$18.801\pm0.070$  & $18.734\pm0.075$ &    $18.704\pm0.031$  \\
     $\sigma_{mean}$  &       $\pm0.017$  & $\pm0.022$ & $\pm0.007$ \\
   \hline 
\end{tabular}
\end{table}

\begin{table}[p]
\centering
\caption{R-band photometric results: mean magnitude and standard deviations. We have removed one of the frames with a FWHM=$1\farcs48$ in the result from the Fitting technique.}
  \begin{tabular}[t]{lrrr}
   \hline
    &\bf Fitting & \bf Clean & \bf Deconv  \\
   \hline
   \bf A & $17.078\pm0.020$  & $17.093\pm0.014$  & $17.106\pm0.021$   \\
     $\sigma_{mean}$   &       $\pm0.006$  & $\pm0.004$ & $\pm0.005$ \\ 
   \bf B & $17.262\pm0.027$  &  $17.274\pm0.014$ & $17.291\pm0.023$ \\
      $\sigma_{mean}$  &       $\pm0.008$  & $\pm0.004$ & $\pm0.006$ \\ 
   \bf C & $18.227\pm0.030$  & $18.109\pm0.032$ & $18.093\pm0.042$     \\
      $\sigma_{mean}$  &       $\pm0.009$  & $\pm0.008$ & $\pm0.011$ \\ 
   \bf D &$18.522\pm0.048$  & $18.441\pm0.039$ &  $18.378\pm0.032$    \\
      $\sigma_{mean}$  &       $\pm0.014$  & $\pm0.010$ & $\pm0.009$ \\ 
   \hline 
\end{tabular}
\end{table}

\begin{table}[p]
\centering
\caption{I-band photometric results: mean magnitude and standard deviations }
  \begin{tabular}[t]{lrrr}
   \hline
    &\bf Fitting & \bf Clean & \bf Deconv  \\
   \hline
   \bf A & $16.957\pm0.027$  & $16.966\pm0.027$  &  $16.961\pm0.025$  \\
     $\sigma_{mean}$  &       $\pm0.006$  & $\pm0.006$ & $\pm0.006$ \\ 
 \bf B & $17.170\pm0.024$  &  $17.192\pm0.030$ &  $17.189\pm0.023$\\
     $\sigma_{mean}$ &       $\pm0.006$  & $\pm0.007$ & $\pm0.005$ \\
   \bf C & $17.973\pm0.054$  & $17.910\pm0.065$ &   $17.976\pm0.042$   \\
     $\sigma_{mean}$ &       $\pm0.013$  & $\pm0.015$ & $\pm0.010$ \\
   \bf D &$18.265\pm0.067$  & $18.206\pm0.083$ &    $18.242\pm0.034$  \\
     $\sigma_{mean}$ &       $\pm0.016$  & $\pm0.020$ & $\pm0.008$ \\
   \hline 
\end{tabular}
\end{table}
\normalsize

\begin{figure*}[ht]
\begin{center} 
\resizebox{8.5cm}{!}{\includegraphics{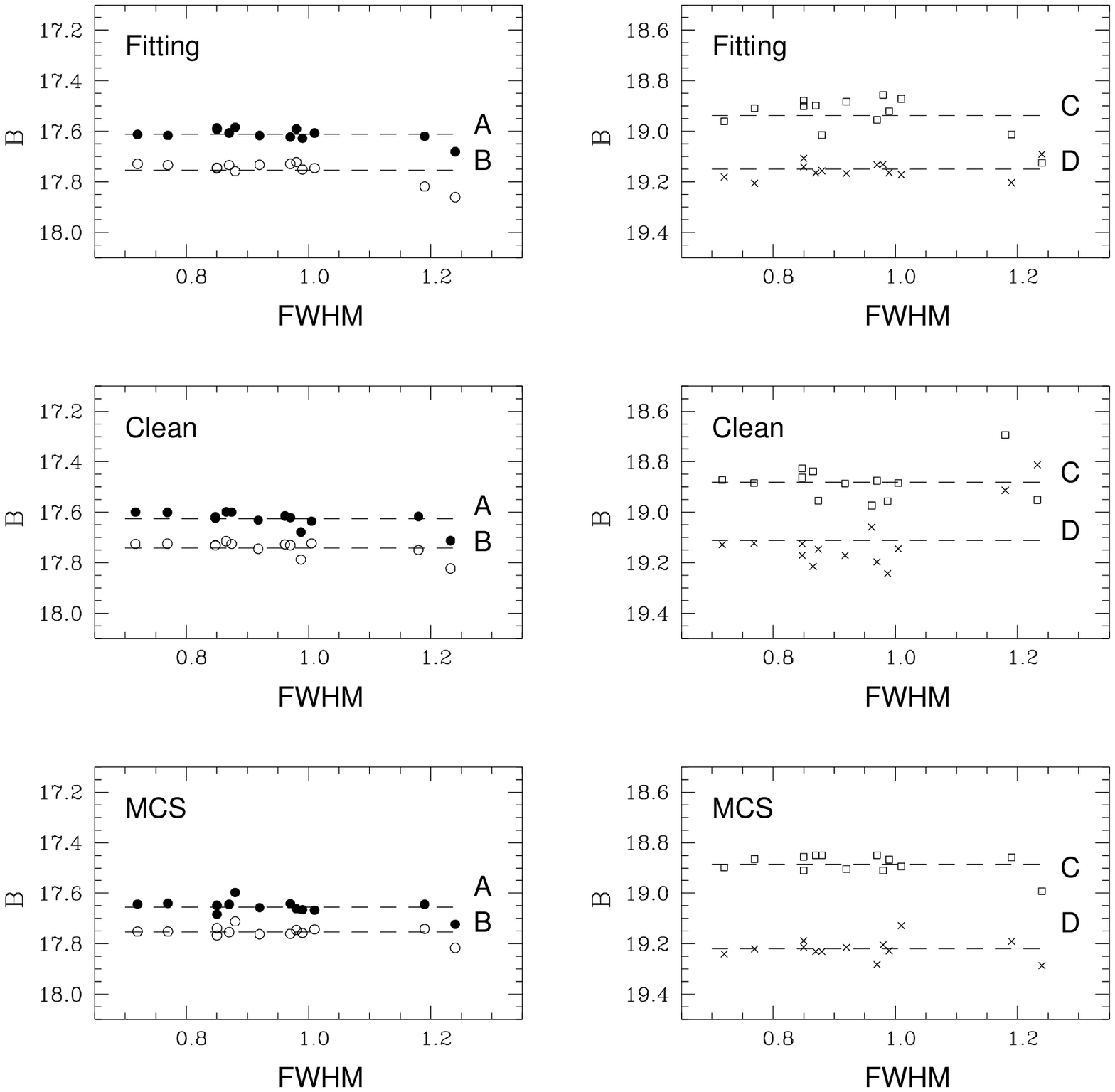}}
\resizebox{8.5cm}{!}{\includegraphics{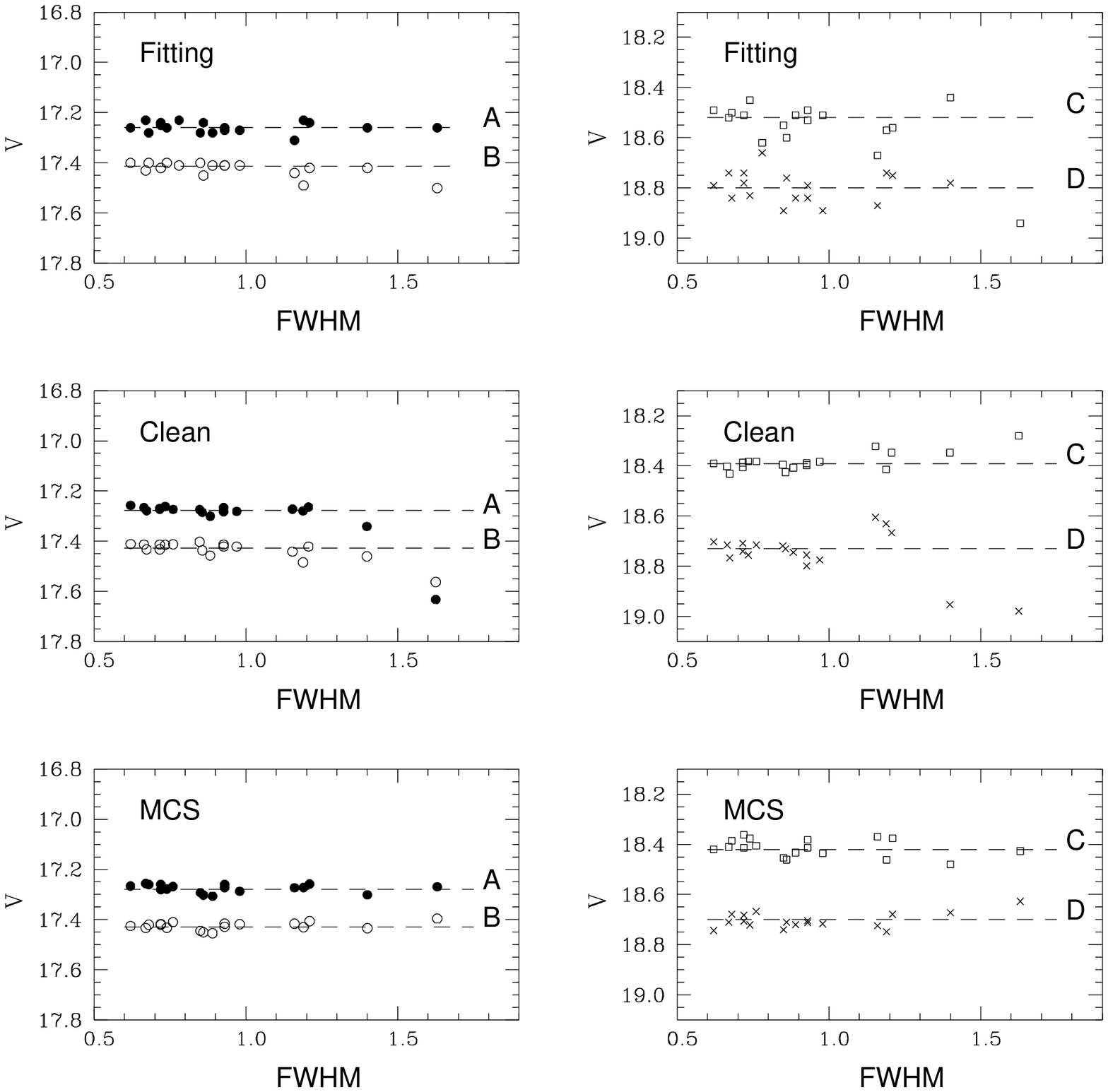}}
\resizebox{8.5cm}{!}{\includegraphics{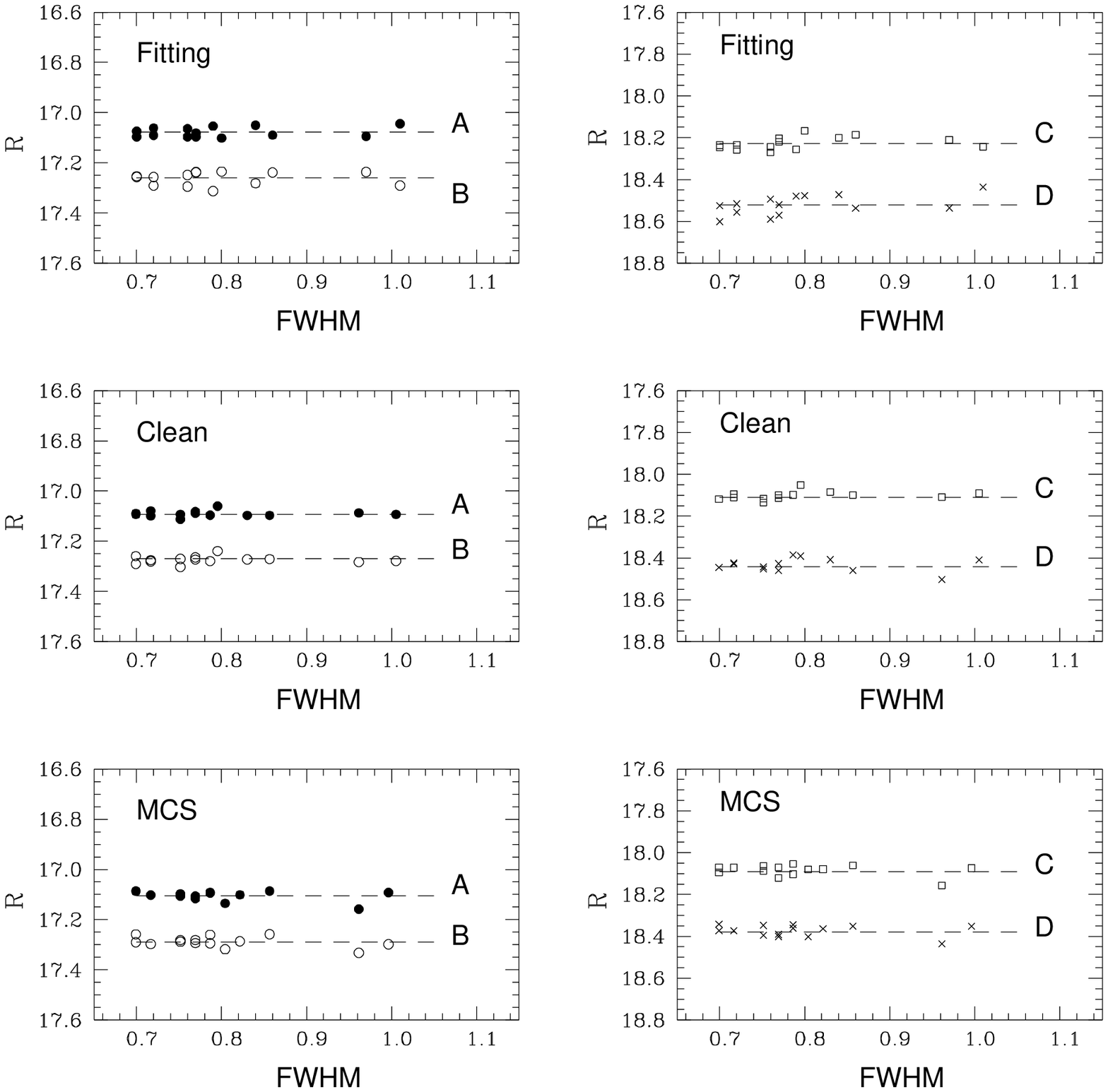}}
\resizebox{8.5cm}{!}{\includegraphics{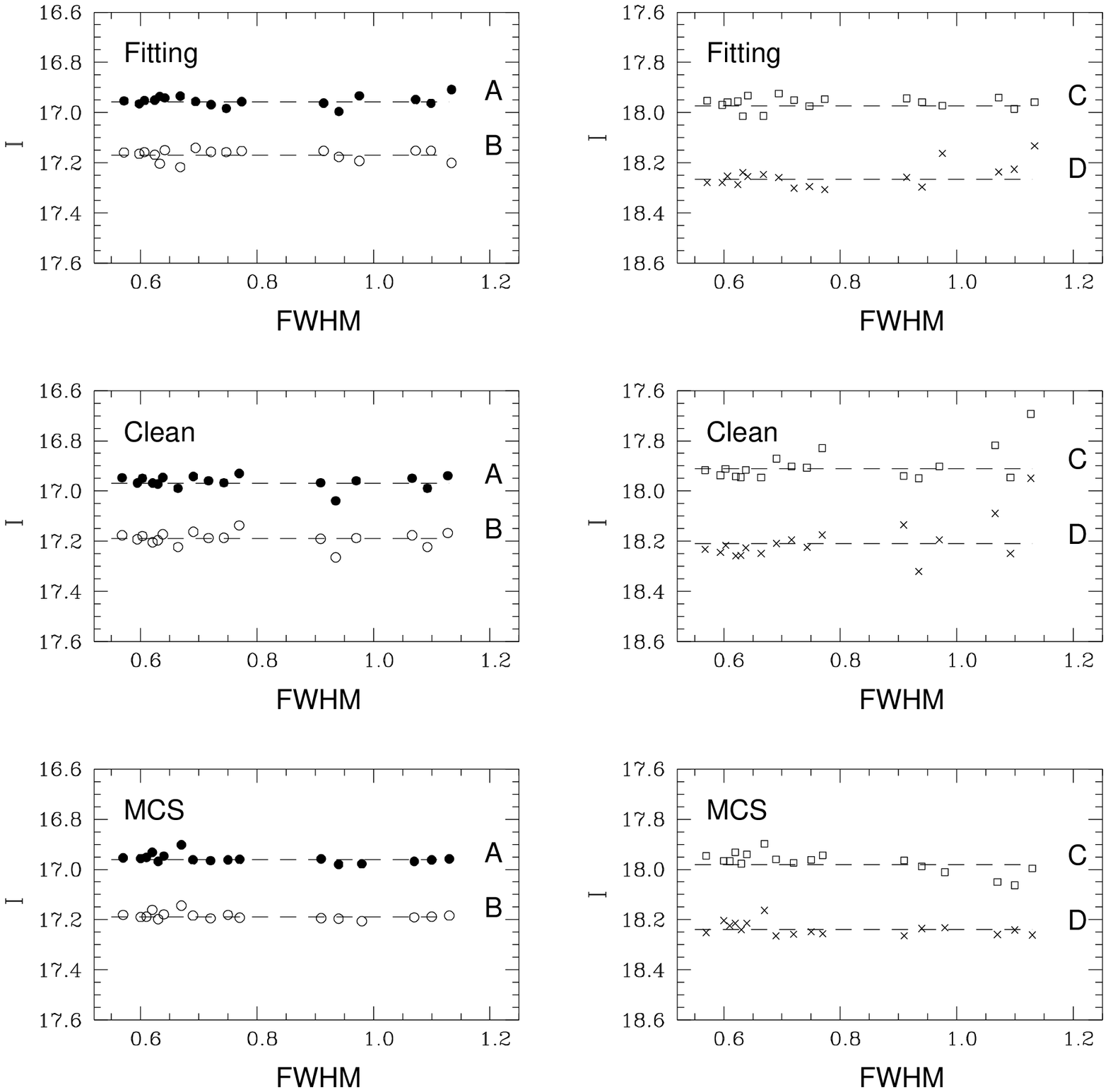}}
\caption{Photometric magnitudes as a function of seeing (in arcseconds)
for the $B$, $V$, $R$ and $I$ frames. Note that in the $V$-band, the
magnitude of component D, derived with the Fitting method on the image with a seeing of 1\farcs63 is outside the plot.}
\label{fig:photometry}         
\end{center}
\end{figure*}

In order  to make sure  that the magnitudes  of the QSO images did not
vary during  the run, we looked  for  possible systematic correlations
between the results of the  different methods. No such correlation was
found.  Therefore we  can safely assume   that the QSO images  did not
vary and we can use the data set to compare the three algorithms.

For the Fitting method and the CLEAN algorithm, the photometry depends
on the seeing of the frame, i.e.,  the errors increase when the seeing
gets worse.  One frame  in the $R$-band  and one in the  $V$-band were
removed from  the analysis using these two  methods,  since they would
have biased the results.  We  find that only the  frames with a seeing
better   than $1\farcs1$  allow  us to   obtain a photometric accuracy
better than 0.05 magnitudes with these two methods.  The MCS algorithm
is less  dependent on seeing  and all the  frames with a  seeing up to
$1\farcs7$   gave photometric  errors below   0.04  magnitudes.  If we
consider data  with an integrated $S/N\sim700$  over one  point source
and a seeing $FWHM\leq1\farcs00$,  the overall $1\sigma$ error  bar is
below  0.02 magnitudes  for the  A  and B  components and  below  0.04
magnitudes  for  the C  and  D  components,  for all   three  methods.
Fig.~\ref{fig:photometry} presents  the magnitudes  for the  $B$, $V$,
$R$   and   $I$ band frames    respectively, measured  with  the three
different techniques.  The magnitudes   are plotted as a   function of
seeing (in   arcseconds).    One of  the   frames  with   a seeing  of
$\sim1\farcs48$ was  discarded from the  $R$-band figures.   We notice
systematic  differences  of the  order of 0.1   magnitudes between the
three methods  for the  C and D  components in  the $R$-band.  This is
more  than the standard deviations  calculated  on the mean values and
indicates that systematic  errors introduced by the analysis algorithm
might significantly  affect the photometry of the  Einstein Cross.  In
several cases, a correlation  between the four  QSO images is observed
in the frames with bad  seeing values.  This  may be due to inaccurate
scaling of the galaxy.
 
The results  displayed  in  Fig.~\ref{fig:photometry}  have  motivated
photometric tests  on simulated   frames  of the object,  in  order to
investigate   the effects  of random   and  systematic  errors  on the
photometry.   Synthetic images were created   with the three different
models of the lensing  galaxy  (de Vaucouleurs profiles and  numerical
models)   and  with  different   seeing  values.   We   found that the
photometric  results from the  Fitting method and  the CLEAN algorithm
depend  on both seeing  and  galaxy  model.  Implementing an  accurate
numerical galaxy model into the Fitting and the Clean algorithms would
therefore  certainly  improve  their  efficiency   and  accuracy,   in
particular for bad seeing  values.  The random  errors are of the same
order  as the statistical  errors we calculated  from the real frames.
The systematic errors are smaller than 0.05 magnitudes for the A and B
component, but up to 0.1 magnitude for the C and D components both for
the Fitting method  and the CLEAN algorithm when  tested on the images
created with  different galaxy models.   No systematic correlation was
found  between the photometry of  images with  different galaxy models
analysed with   different methods, indicating   that these  systematic
errors are difficult to correct for.

For the deconvolution method,  the systematic errors are  smaller than
0.05   magnitude for all the  components,  and the results depend much
less on the  seeing  of the frames  since a  detailed numerical galaxy
model is used.

\subsection{Astrometric results}

\begin{figure}[t]
\begin{center}
\resizebox{8.5cm}{!}{\includegraphics{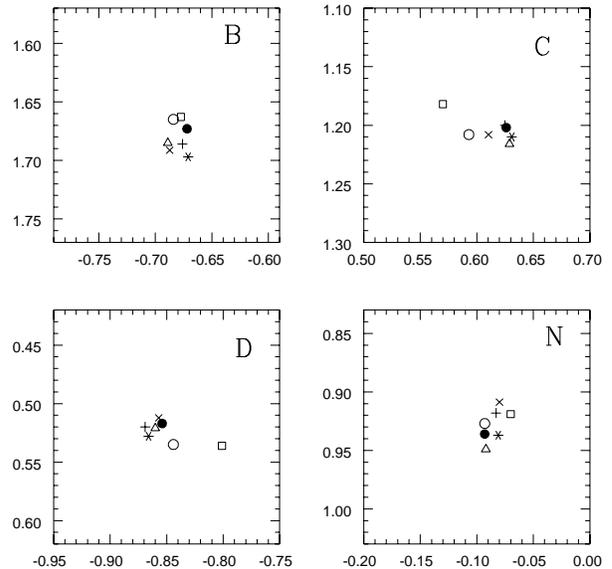}}
\caption{Mean   positions   relative to  the    A component   from all
methods.  The $x$ and $y$ axis  give respectively R.A. and Declination
in arcseconds relative to A.  The symbols correspond to the following:
$\times$  Fitting, $\Box$ CLEAN,  $\triangle$  MCS, $\bullet$ Crane et
al. , + Rix et al., $\circ$ {\O}stensen et al., $\ast$ Blanton et al.}
\label{fig:comp}
\end{center}
\end{figure}

The  relative positions  measured  for the four   components using the
different methods are  in fairly  good agreement,  not  only with  one
another, but also with the positions determined from observations with
the HST (Crane et al. \cite{crane}, Rix  et al. \cite{rix} and Blanton
et   al.  \cite{blanton})   and  the   results  from the   photometric
monitoring  published    by   {\O}stensen   et   al.    (\cite{royb}).
Fig.~\ref{fig:comp} presents the geometry   of the object relative  to
component A,   as derived from the  three  methods,  and compared with
previously published astrometry.   Note that  only  the frames with  a
sub-arcsecond seeing were used for   the Fitting method and the  CLEAN
algorithm.   We found  that  the  astrometric  accuracy of  these  two
methods significantly decreases with  bad seeing.  With a seeing worse
than  $1\farcs1$,  typical errors  of  $0\farcs02$  are observed.  The
errors seem  to be random with   the Fitting method,  whereas with the
CLEAN algorithm, the positions measured for the C and D component tend
to drift towards the galaxy's nucleus when the seeing gets worse.

The astrometry obtained with  the deconvolution algorithm is much less
seeing  dependent  as soon as  an  accurate numerical galaxy  model is
used.  However,  obtaining such a galaxy  model  requires the nucleus'
position  to  be  well  estimated  from  high   S/N data  (see Section
3.3). This  is particularly true in  the $B$-band where the nucleus is
very faint and its position less accurate (about $0\farcs05$).

\section{The lensing galaxy}

\subsection{Morphology}

\begin{figure}[t]
\begin{center}
\resizebox{9.0cm}{!}{\includegraphics{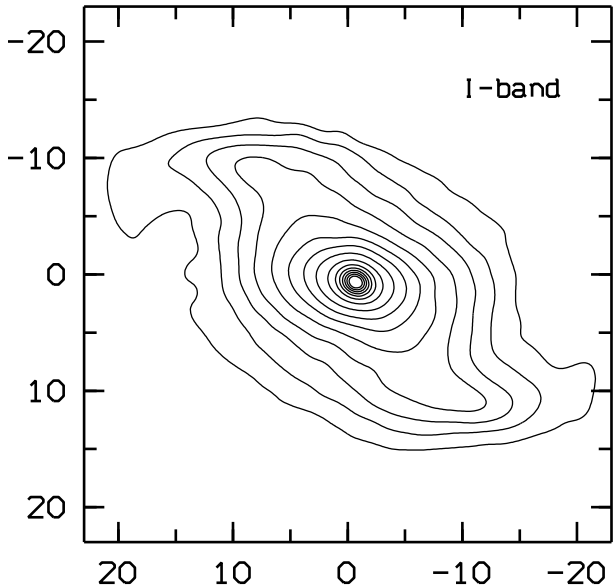}}
\caption{Contour plots of the numerical galaxy model from the MCS
deconvolution code in the $I$ band. Both axis are in arcseconds.
North is up and East is to the left.}
\label{fig:galcont}
\end{center}
\end{figure}

As described in Section 3.3 and  shown in Fig.~\ref{fig:rima}, the MCS
code  produces a deconvolved numerical   galaxy both with and  without
nucleus.  The contour plot displayed in Fig.~\ref{fig:galcont} clearly
reveals a bar in  the central 10 arcseconds  of  the galaxy. It  has a
very similar shape  in the $V$  and $R$ bands  whereas in the $B$ band
the S/N of the data  is too low to derive  an accurate galaxy profile.
The bar cannot be correctly modeled by a pure analytical profile, even
in the centre of  the galaxy, as in fact  suggested by the significant
residuals   obtained    with   the   Fitting      and    the     Clean
algorithms.  Photometry obtained by   using only analytical  models is
therefore likely to be biased.

\subsection{Extinction}

\begin{figure}[t]
\begin{center}
\resizebox{8.5cm}{!}{\includegraphics{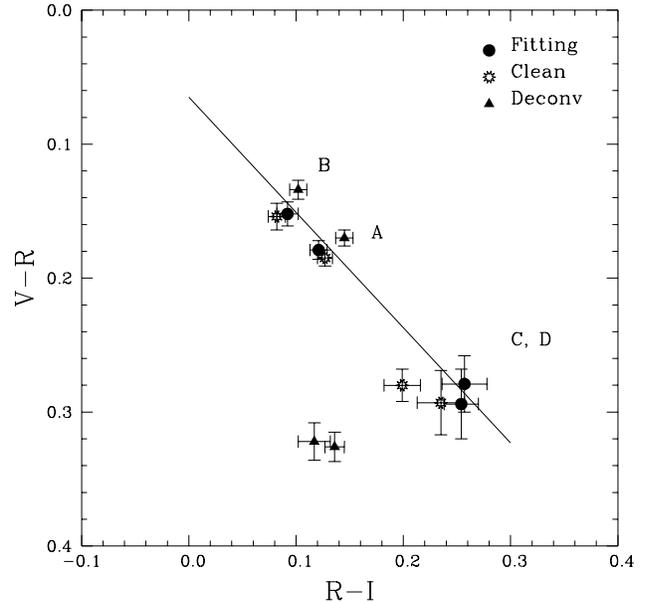}}
\caption{V-R as function of R-I for the three different methods with error-bars
on the mean of all the frames ($\sigma_{mean}$ in Tables 2-5).
The mean extinction law for our Galaxy is drawn as a solid line through
the mean colour of the A component. See Section 5 for an explanation of
the C and D points.}
\label{fig:colour}
\end{center}
\end{figure}

From our analysis, we found that the A  component is reddened relative
to B, and that the C and D components are both reddened relative to A.
Fig.~\ref{fig:colour}  shows the $V-R$  colour as a function of $R-I$.
The error-bars  indicated are the errors  on the mean ($\sigma_{mean}$
in  Tables 2-5).    If the reddening   is  due to extinction  from the
galaxy, the four quasar images should  lie approximately on a straight
line representing  the extinction law for the   lens galaxy.  Since it
has been suggested  that this law is similar  to the one of our Galaxy
(Yee \cite{yee}),  the mean extinction law for  the latter is shown on
Fig.~\ref{fig:colour},  as a solid line with  a slope of 0.86 (Vakulik
et  al.   \cite{vakulik}).  Cardelli et  al.  (\cite{cardelli}) showed
that  in the  $V$-$R$-$I$ domain, the  shape of  the extinction law is
insensitive to  the parameter  $\rm  R_{\rm V}=\rm  A_{\rm V}$/E(B-V),
thus independent on the  origin of the extinction (e.g.,  interstellar
matter, dense clouds).   From Fig.~\ref{fig:colour},  the results from
the Fitting and the CLEAN  as well as  the A and  B magnitude from MCS
could be interpreted as extinction.  The positions of components C and
D  compared to A   and B  on the  straight   line can be explained  by
reddening due to the tilt of  the galaxy in the direction  of C and D.
This  tilt might   even explain the  reddening  of  A  compared  to B.
However,  we point out  that the magnitudes  obtained with Fitting and
CLEAN  in the different   bands do not  agree  with each other, but by
coincidence the differences cancel out in this particular colour plot.

Considering the  differences in C and  D between  the three methods we
can not confirm the presence of  a mean extinction law. In particular,
the C and D  results  from MCS could be  interpreted  as a long   time
colour   dependent microlensing effect.  When    the  source passes  a
microcaustic, its inner  and hence bluer parts  may be  more amplified
than the outer emission line region of the accretion disk.  (Kayser et
al.  \cite{kaysera} and Wambsganss \& Paczy{\`n}ski, \cite{wambsganss}
).  This is  even true for sources  angularly larger than the Einstein
radius of the lens (Refsdal \& Stabell \cite{refsdal}).

We  found a   colour  excess of   B  with respect   to A   of  $\Delta
E(V-I)=-0.08\pm0.02$, magnitudes. This is in agreement with Vakulik et
al.    (\cite{vakulik})     who  found   $\Delta  E(V-I)=-0.12\pm0.05$
magnitudes, also from observations taken in 1995.  However, Yee et al.
(\cite{yee})     found   that   B   was    reddened   relative   to A,
$E(g-i)=+0.08\pm0.03$.

A possible change in the relative colours over time might be explained
by colour dependent microlensing or,  less likely, by dust  extinction
that varies over time (e.g., Rix  et al.  \cite{rix}).  The reason for
a colour change in the Einstein  Cross will remain unclear unless long
term     multicolour  photometric   monitoring     (   or   preferably
spectrophotometry) is carried out on a dedicated telescope.

\section{Discussion and conclusions}

The two most complete light curves from  photometric monitoring of the
Einstein        Cross  have  been     published      by Corrigan    et
al. (\cite{corrigan})  and  more    recently by {\O}stensen    et  al.
(\cite{royb}).  Although several obvious microlensing events have been
observed,  no  precise estimates  of the photometric  errors have been
discussed so  far. Moreover, better sampling  of the  light curves, as
well as more accurate photometric measurements are  needed in order to
interpret  the  observed  intensity  differences  in  detail.   In the
present study, a set of observations in four bands was obtained during
two successive nights.  Since the time delay for the Einstein Cross is
of the order of one day, observations on a short time scale would make
it   possible  to  separate  intrinsic   variations  from microlensing
effects.

A bright star is available on all our frames to perform PSF photometry
with the   three different methods considered  in   this paper.  The
seeing  during the   two nights was   variable ($0\farcs6$-$1\farcs7$),
which  allowed us to quantify  the seeing  dependence of the different
methods applied.

Differences are observed between the photometric measurements obtained
with the  three methods    (see Fig.~\ref{fig:photometry}).   All  the
methods show  random errors that, as  one should expect, increase with
bad seeing.   In addition,  systematic errors in  one or  more  of the
methods  are also present,  especially for  the fainter components  of
Q~2237+0305.  For each QSO  component, the dispersion between the mean
magnitude obtained with   the 3 techniques is  larger   than the error
measured ($\sigma_{mean}$) for  each method.  Photometric measurements
performed on simulated frames also  showed that the systematic  errors
depend both on the seeing and on the galaxy model used to describe the
deflector.  The Fitting and the  Clean algorithms produce large errors
(0.04-0.1  magnitudes) for the two  faint quasar  components (C and D)
when  the seeing  is worse than  $1\farcs0$.   The errors from the MCS
algorithm are  below 0.04 magnitudes and  less dependent on the seeing
when a numerical galaxy model is used.

If only subarcsecond  seeing frames are  considered, we find that  all
three methods could have  detected intensity variations with a minimum
amplitude of  0.02 and 0.04  magnitudes in  components A, B  and C, D
respectively.     The  latter errors are     given for one individual
measurement and  can be improved by using   several frames.  When the
exposures are  of $\sim$200s, as  in our case,  about 10 frames can be
obtained in  one   hour   of observation.   If   10   frames with   a
subarcsecond   seeing  are  obtained,    the   random error will    be
$0.02/\sqrt{10}=  0.006$ and $0.04/\sqrt{10}=0.013$ magnitudes for A,
B and C, D respectively.

Assuming that systematic errors are present, it is reasonable to think
that they  would be more important  when an analytical galaxy model is
used, rather than when a  purely numerical model  is applied.  Even if
an analytical de Vaucouleurs profile can fit the  central bulge of the
galaxy, it cannot  model the bar in the  spiral galaxy,  and we cannot
exclude  that  this may affect  the  photometry of the  quasar images.
Moreover, a  good  numerical model   obtained from high   S/N data can
represent  high   frequency structures in the    galaxy which would be
neglected by an analytical model.  For example, H\,{\sc ii} regions in
the  lens galaxy  can only be  modeled by  a  numerical profile.  Such
clumpy structures, likely to emit in the $H_{\alpha}$ line, would then
make inaccurate the  $R$-band photometry obtained with  any analytical
galaxy  model.  In such   a  case, the photometry  of   the  C and   D
components would be most affected since  their positions are along the
bar of the galaxy, where most  of the H\,{\sc  ii} regions should lie.
The largest  photometric differences between  the methods investigated
here  are indeed   present in the   $R$-band, but $H_{\alpha}$-imaging
would be  needed  to confirm such an  effect  by directly imaging  the
H\,{\sc ii} clumps.  We should also point out that finding a numerical
galaxy involves a larger number   of free parameters.  In the  present
case, we cannot  rule out the possibility  that the S/N in our  frames
was too low to obtain a unique numerical model.

Based on the published light curves and  the present study, we propose
the  following   observational strategy   for   a future   photometric
monitoring of the Einstein Cross.  The aimed sampling of the points on
a light  curve  should  be  compared with   typical   time scales   of
microlensing events.  Since time scales as short as  14 days have been
observed ({\O}stensen et al.  \cite{royb}), a sampling of the order of
a  few observations per night  should be aimed at.  Furthermore, since
intensity   variations of 0.05   magnitudes  or less  can be  expected
({\O}stensen et al.  \cite{royb}),   it  is  important to  ensure    a
photometric accuracy  well below this value.   As shown in  our study,
using  several data frames   (up to 10 with   a 2m class telescope) to
derive any individual photometric measurement would yield random error
bars   smaller   than 0.01    magnitude   for  subarcsecond  observing
conditions, and would allow such a monitoring program.

However, one cannot expect subarcsecond conditions over observing runs
as long as several nights.  Given this observational constraint, it is
likely that the best photometric accuracy will  be obtained by using a
detailed   numerical galaxy model. We  have  shown  that such a galaxy
model can be  constructed from high S/N  frames with the MCS algorithm
and that accurate photometry can be achieved by applying this model in
connection with simultaneous deconvolution of all the frames.

\begin{acknowledgements}
We wish to  thank   S.  Sohy for  help   with the MCS  algorithm,  and
J.P. Swings for support.   IB, FC, and R{\O} are  supported in part by
contract    ARC94/99-178 ``Action de   Recherche   Concert{\'e}e  de la
Communaut{\'e}  Fran\c{c}aise    (Belgium)''  and  P{\^o}le   d'Attraction
Interuniversitaire, P4/05  (SSTC, Belgium).

\end{acknowledgements}
\end{document}